\documentclass[twocolumn]{ceurart}

\usepackage[utf8]{inputenc}
\usepackage{amsmath}  
\usepackage{amsfonts} 
\usepackage{graphicx} 
\usepackage{physics}
\usepackage{comment}
\usepackage{url}
\usepackage[labelfont=bf]{caption}
\usepackage[a4paper, total={6.5in, 9in}]{geometry}
\usepackage{pdfpages}

\begin{document}

\title{Endless Fun in high dimensions - A Quantum Card Game}

\author[1]{Lea Kopf}
\address[1]{Tampere University, Photonics Laboratory, Physics Unit, Tampere, FI-33720, Finland}
\author[1]{Markus Hiekkam\"aki}
\author[1,2]{Shashi Prabhakar}
\address[2]{Quantum Science and Technology Laboratory, Physical Research Laboratory, Ahmedabad, India 380009}
\author[1]{Robert Fickler}

\begin{abstract}
Quantum technologies, i.e., technologies benefiting from the features of quantum physics such as objective randomness, superposition, and entanglement, have enabled an entirely different way of distributing and processing information. 
The enormous progress over the last decades has also led to an urgent need for young professionals and new educational programs. 
Here, we present a strategic card game in which the building blocks of a quantum computer can be experienced. 
While playing, participants start with the lowest quantum state, play cards to ``program'' a quantum computer, and aim to achieve the highest possible quantum state.
Thereby they experience quantum features such as superposition, interference, and entanglement. 
By also including high-dimensional quantum states, i.e., systems that can take more than two possible values, and by developing different multi-player modes, the game can help the players to understand  complex quantum state operations and can also be used as an introduction to quantum computational tasks for students. 
As such, it can also be used in a classroom environment to increase the conceptual understanding, interest, and motivation of a student. 
Therefore, the presented game contributes to the ongoing efforts on gamifying quantum physics education with a particular focus on the counter-intuitive features which quantum computing is based on.
\end{abstract}
\maketitle

\section{Introduction}
Quantum physics is considered one of the most successful branches in physics that humans have conceived so far. 
At the very heart of quantum physics are principles such as objective randomness, interference, superposition, and entanglement. 
These concepts are difficult to grasp as they often contradict our intuition, which is based on everyday experiences and a classical understanding of the world. 
It is these counter-intuitive features that have enabled novel technologies which would not be possible in a classical setting.
The current technological thrive is often termed as the second quantum revolution \cite{dowling2003quantum}, since it differs significantly on the fundamental and applicational level from the already established technologies based on quantum physics. 
One of the most prominent examples of quantum technology is a quantum computer. 
Compared to their classical counterpart, quantum computers promise a speed-up of certain tasks, outperforming all modern computers and enabling computational algorithms that are impossible in a classical setting. 
Recent progress has led to the first large-scale quantum computational system that outperforms all classical computers in a specific computational task \cite{arute2019quantum, zhong2020quantum}. 

In this article, we present a card game that uses gamification strategies to provide a fun and engaging introduction to the concepts of quantum computing. 
The game, \emph{Endless Fun in high dimension}, is designed to be a low threshold introduction to quantum computing and encourage people to look into the fundamentals of quantum physics. 
The game implements basic quantum mechanical concepts such as quantum logic operations, superposition, and entanglement. 
While playing cards which correspond to quantum operations, the participants are manipulating the high-dimensional quantum state of a quantum computer with the aim of achieving the highest value of their own quantum state. 
The final result is evaluated using an included computer program. 
We note that although current quantum computers work with two-level quantum states, the presented card game goes a step further by also including the programming of high-dimensional quantum system.
While such states are considered promising candidates for next generation quantum computers and other technologies \cite{cozzolino2019high, wang2020qudits, erhard2020advances}, they are also beneficial in the presented gaming setting as they increase the complexity of the game play and as such enable longer lasting (maybe endless) fun.

\section{Background}

As deeper understanding of a new concept often starts with an initial intuitive grasp of the effects, a gameful approach to complex topics, e.g., quantum physics and quantum information, has been the focus of various gamification efforts \cite{deterding2011game, hamari2016challenging}.
They offer an entertaining way to loosen up the atmosphere in a course, to enhance understanding, and to promote academic findings \cite{kuo2016gamification}.
Education in quantum physics and, in particular, modern quantum information science can benefit from ideas developed through gamification methods.

\subsection{Gamification in quantum physics education}

A popular approach to quantum physics education is to demonstrate various quantum effects in computer simulations, e.g., videos that visualize quantum effects \cite{thaller2005advanced}, or to get an hand-on experience when using adjustable experimental setups \cite{Quantenkoffer}.
Quantum games, which cover similar effects, are focusing more on gamification methods, such as the online computer game \emph{Particle in a Box} \cite{anupam2017particle}, or ``quantized'' adaptations of well-known games: 
\emph{Quantum TiqTaqToe} \cite{goff2002quantum, QTicTacToe}, \emph{Quantum Chess} \cite{QuantumChess,cantwell2019quantum}, \emph{Quantum Minigolf} \cite{QuantumMinigolf}, as well as the quantum version of Minecraft \emph{qCraft} \cite{QCraft}.
An extensive overview over the increasing number of quantum games with a focus on quantum computer games can be found in \cite{piispanen2022games}.
In addition to online games, there are also educational board games, e.g., \emph{Entanglion} by IBM \cite{Entanglion}.
The idea of gamification in quantum sciences has also been the center of focus in various quantum game jams \cite{QuantumWheel}, in which instructive and entertaining games have been developed. 
Many quantum physicists hope that these games could be more than just tools for learning, as expressed by John Preskill who states that ``[p]erhaps kids who grow up playing quantum games will acquire a visceral understanding of quantum phenomena that our generation lacks''  \cite{preskill2018quantum}.

\subsection{Related work - \emph{Q$\ket{\text{Cards}}$}}
At the Quantum Wheel game jam in Helsinki in 2019 \cite{QuantumWheel}, the quantum card game \emph{Q$\ket{\text{Cards}}$} was developed and introduced \cite{QCards}.
The game presented here builds upon and extends \emph{Q$\ket{\text{Cards}}$}.
The gameplay is very similar and only varies in minor details. 
However, in contrast to \emph{Q$\ket{\text{Cards}}$}, \emph{Endless Fun} aims at building a high-di\-men\-sio\-nal quantum computer, i.e., a quantum computer operating on states with three possible values instead of two. 
Although current quantum computers exclusively use binary-valued quantum systems to encode bit-valued quantum information, the increase in possible outcomes when using high-di\-men\-sio\-nal quantum systems further enhances the complexity of the game and offers more options to adjust the difficulty. 
Thus, we anticipate a longer-lasting interest in playing the game. 
Furthermore, the game includes a cooperative and a single-player game mode in which different learning objectives are addressed.
To allow a simple determination of the game's outcome, Python codes based on the mathematical framework needed in higher dimensions were exclusively developed for the card game.

\section{Endless Fun in high dimensions}
\emph{Endless Fun} is a strategic multi-player card game that introduces the players to quantum computational logic gates. 
By playing quantum gates, the players aim to increase their quantum state value and decrease the quantum state values of the other players. 
The final quantum state values are calculated by an evaluation software. 
Comparing the expected outcome to the mathematically correct outcome, provided by the software after a card is played, allows one to reflect on misunderstandings and retrace the effects of each operation. 
The main goal of this game is to provide a platform to practice and engage with quantum logic operations while providing varying difficulty levels.
The detailed instruction manual, the Python-based evaluation software, the cards, and the riddles of the single-player game mode are provided as supplementary material and in an online repository \cite{kopf2020Endless}.

\subsection{Game concept}

\begin{figure*}[ht]
  \centering
  \includegraphics[width=\linewidth]{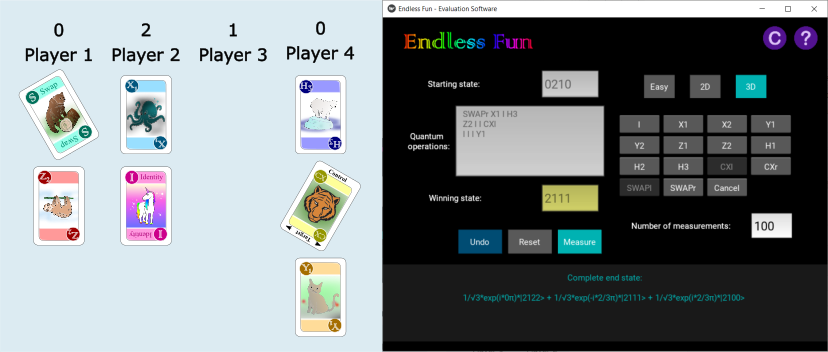}
  \caption{A possible second-round with 3-dimensional qudits and four players. 
  Player 1 starts with the qudit value of $\ket{0}$ obtained in the first round.
  Player 2, 3, and 4, start with their qudit values of $\ket{2}$, $\ket{1}$, and $\ket{0}$, respectively.
  The software on the right evaluates the winning state as $\ket{2,1,1,1}$, where the position of qudit values corresponds to the players. 
  For a better understanding of the underlying mechanisms, the generated end state is also displayed on the bottom of the window, before it was measured.
  The other possible outcomes can be found by inspecting the overall state. Here the states $\ket{2,1,2,2}$ and $\ket{2,1,0,0}$ could also have been obtained with the same probability.}
  \label{fig2}
\end{figure*}

In classical computation, information is saved in bit values, which can either be 0 or 1. 
Analog to classical computing, binary or 2-dimensional (2d) quantum computers encode information into two-level quantum systems, so-called quantum bits or \emph{qubits}. 
Similar to classical bits, qubits can take values of 0 or 1. 
However, the quantum nature also allows for superpositions thereof, i.e. loosely speaking being in both states at the same time. 
To make this difference also visually clear, qubit states are commonly written in the abstract bra-ket-notation introduced by Dirac, i.e., $\ket{0}$ or $\ket{1}$. 
In this notation, any superposition is written as a sum over both possible states, e.g., $\frac{1}{\sqrt{2}}(\ket{0}+\ket{1})$, which means that any measurement can result in either outcome, 0 or 1, with equal probability. 
Note that the prefactor $\frac{1}{\sqrt{2}}$ is a measure for the probability amplitude, which is the square root of the probability \cite{audretsch2008entangled}. 
High-dimensional quantum states, often called \emph{qudits}, go a step further in complexity, as they do not only allow two but \emph{d} possible outcomes. 
The general goal of the game is to perform quantum state operations that are building blocks to program a quantum computer by playing the cards in such a way that the player's own qudit value is as high as possible and the opponent's values as low as possible.
An example game-play is shown in Fig.~\ref{fig2}.
The game can also be played in a cooperative mode or single-player game mode.
In the cooperative mode, the goal is not to win against the other players but to reach the highest possible values summed up over all qudits as a group. 
In the single-player mode, the player can solve six ready-made riddles which guide the player to discover specific quantum effects.
The riddles have different levels of difficulty, starting with easy ones that help the player learn about quantum interference effects.
The difficulty is then gradually increased, with more quantum effects being gradually introduced.
In addition, the significance of each quantum effect in quantum computing is briefly discussed along with the solution of the riddle, such that students can put the learned quantum operation in a better context. 
Instructors can extend this set with their own riddles.
For a detailed description, see the rules of the game in the instruction manual.

\subsection{Quantum operations in 2d and 3d}
The game can be played with either 2- or 3-dimen\-sio\-nal quantum logic operations. 
In 2d, only the qubit values $\ket{0}$ and $\ket{1}$ are available, in 3d an additionally qudit value $\ket{2}$ is available.
The choice of dimension also affects the number of quantum operations. 
In 2d, for example, only one X-gate is defined, whereas in 3d, two X-gates with different behaviors exist. 
To increase possible winning strategies the game also includes cards, which are not corresponding to quantum operations, e.g. a steal card that allows a player to steal a card from another player.
The instruction manual gives an overview of all playable operations and the detailed truth tables for all quantum gates.
A good winning strategy is to keep track on the evolution of the state.

To not overwhelm the players that do not have a solid background in quantum information, it is recommended to start the game in the \textit{Easy} version. 
In this simplified 2d-version, the beginners can familiarize themselves with the rules and basic quantum logic operations without the phase properties of the operations. 
The game is more complex in the standard \textit{2d} version, in which cards are added that modulate the phase of single states which allows the players to control quantum interference. 
Finally, the \textit{3d} version is played with three-dimensional qudits, thus it includes the most complex quantum states and the gameplay reaches its maximal difficulty.
More details on the exact set of cards used in each version can be found in the manual.

\subsection{Quantum effects}
In the game, three quantum effects can be investigated:
\begin{itemize}
    \item Quantum superpositions, which demonstrate the probabilistic nature of quantum measurements.
    \item Quantum interference, which demonstrates the effect of phases on measurement outcomes.
    \item Quantum entanglement, which leads to strong correlations between the measurement outcomes of different quantum systems.
\end{itemize}
In the following, we give simple examples how the three effects can be observed in the card game.
For simplicity, we explain the effect in detail with qubits, however, the high-dimensional counterparts follow in an analogous manner.

\begin{figure*}[ht]
  \centering
  \includegraphics[width=0.65\linewidth]{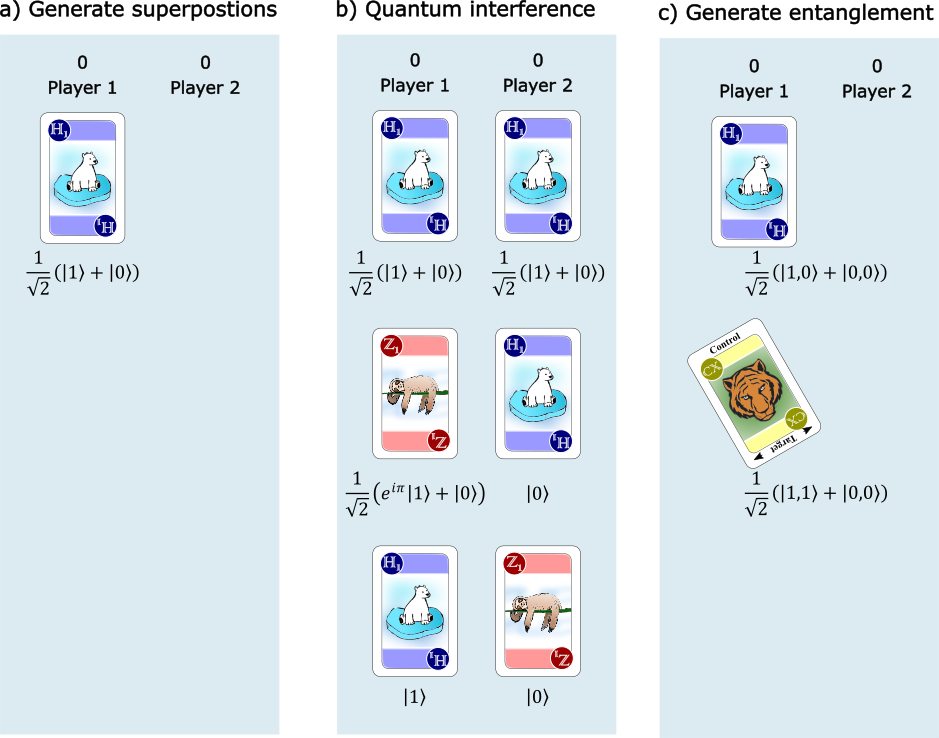}
  \caption{Generating superpositions, interference, and entanglement with qubits. The evolution of the quantum states is shown beneath the played quantum operation. a) A quantum operation is generating a qubit superposition of the state of player 1. 
  Player 1's starting state $\ket{0}$ turns into $\frac{1}{\sqrt{2}} (\ket{1}+\ket{0})$. 
  b) Quantum interference can be controlled by phase. 
  Player 1's value after the logic operations is $\ket{1}$, Player 2's is $\ket{0}$. 
  c) Entangling the qubits of players 1 and 2. 
  The starting state $\ket{0,0}$ is transformed to the entangled state $\frac{1}{\sqrt{2}}(\ket{1,1}+ \ket{0,0})$. 
  The qubit values of player 1 and 2 could be any one of the two possible states, but the values of the two players will always match perfectly.}
   \label{fig3}
\end{figure*}

\paragraph{Quantum superposition}
A Hadamard gate acting on the quantum states $\ket{0}$ or $\ket{1}$ generates a superposition of both states, as shown in Fig.~\ref{fig3} a). 
By playing a Hadamard gate on a qubit, the player's value is, loosely speaking, in $\ket{0}$, and $\ket{1}$ at the same time. 
Only when observing the state in a measurement, it takes on the value of $\ket{0}$, or $\ket{1}$, such that it can be found in either state with the same probability. 
In the game, the software ``collapses'' the state by simulating a measurement and gives the random outcome with the correct quantum probability. 
This probability adds an element of luck to the gameplay, especially if only a small number of measurements is used in the evaluation program to obtain the final state. 
Additionally, the superpositions can be used as a strategic element.
For example, if one of the players is leading the round, the others can set this player into a superposition, reducing their changes of winning.
When the game is played in the 3d version, similar Hadamard operations can be performed, however, with the superposition having three possible outcomes.
Note that the variety of Hadamard gates only differ from each other in phase (see manual for more details), which only become important when considering interference effects.

\paragraph{Interference effects}
The phase of a quantum state is a physical property which does not have a direct effect on its qubit values. 
However, as it affects the outcome of quantum interference it can indirectly be used to change the qubit value of a state.
When playing the game, it is possible to learn how to control interference through phase manipulations. 
Controlling phase is an important underlying working principle of quantum computations and almost always the reason behind its quantum advantage.
In a simple example shown in Fig.~\ref{fig3} b), we assume that we have a qubit $\ket{0}$ on which we play two H$_1$-Hadamard operations. 
In this process the second Hadamard allows interference to occur, resulting in the state $\ket{0}$.
If we add a Z-gate before or after the two H$_1$-gates, we get $-\ket{0}$, which is still the $\ket{0}$-state but with a (global) phase factor that is not relevant for the outcome of the game when the state is measured. 
If however, we first play the H$_1$ gate, then the Z-gate, and then the second H$_1$-gate, the resulting state is $\ket{1}$.
A quick look at the state evolution shows that after the first card we obtained the superposition state $\frac{1}{\sqrt{2}}(\ket{0}+\ket{1})$. 
The Z-gate then changes the phase between the two terms, i.e. changing the state to $\frac{1}{\sqrt{2}}(\ket{0}-\ket{1})$, which leads the final state $\ket{1}$, when another H$_1$-gate is applied. 
Thus, with phase we can manipulate the evolution of a superposition to obtain a desired state, thereby controlling the probability of measuring it, which is also known as quantum interference.
The phase gates can thus be used to control the state, and the measurement outcome, through quantum interference.
Interference effects can also be observed when the game is played in 3d, where the increased complexity of the states allows a larger variety of different phase manipulations and interference effects.
In this game, the Y, Z, and Hadamard cards can be used to change the phase of a quantum state.
Interference effects can be explored in a guided manner in the two easy riddles of the single-player game mode.

\paragraph{Quantum entanglement}
Entanglement is an\-other fundamental feature of quantum mechanics. 
Quantum entanglement correlates the value of one qubit with the value of another qubit. 
Counter-intuitively, the correlation of entangled qubits still exist in multiple states simultaneously. 
Hence, when an entangled state is measured, the outcome of entangled players will be random due to being in a superposition, but still perfectly correlated. 
By using quantum entanglement in the game, you can for example ensure that a certain opponent does not get more points than you.
A plethora of other possible strategies open up when considering tuning the correlation through other gates, e.g. phase gates. 

In a quantum computation process and, thus, in the game, entanglement is generated by playing a Hadamard-gate and consecutively a CX-gate on one qubit. 
As an example, let’s assume players 1 and 2 both have a qubit value of $\ket{0}$, as displayed in Fig.~\ref{fig3}~c). 
A Hadamard-card is played on qubit 1, generating a superposition, i.e., the two-qubit state becomes $\frac{1}{\sqrt{2}}(\ket{0,0}+\ket{1,0})$. 
If we then play a CX-gate (controlled by qubit 1 while targeting qubit 2), the resulting state is $\frac{1}{\sqrt{2}}(\ket{0,0}+\ket{1,1})$. 
This means, that both players' states will randomly have either the value 0, or 1 after a measurement is performed. 
However, due to entanglement, both qubits will always end up with the same random value. 
The same entangling operation also works in 3d.
In the single-player mode, the player is guided through instructive examples of entanglement in three different riddles with varying difficulty.

\subsection{\emph{Endless Fun} in higher education}

In the authors' experience, people who have not encountered quantum mechanics in their studies are generally reluctant to approach quantum computation, since it is perceived to be a highly sophisticated and demanding topic. 
To encourage active participation and get past feelings of reluctance when dealing with the topic, the \emph{Endless Fun} card game offers an easy approach to forming a conceptual understanding of the topic.
By comparing the expected state when cards are played with the evolved states, displayed in the provided computer code, will increase the understanding of the complex quantum operations without the necessity to fully understand the underlying mathematical framework.
For more experienced players, the game offers an environment to apply and extend their knowledge and reflect on misunderstandings. 
In the frame of quantum information and computation courses, playing the game can have additional advantages. 
It adds diversity to classical teaching methods, prevents boredom, and motivates students through positive feedback \cite{Vlachopoulos2017effect}.
Furthermore, it gives room to experience, explore, and practice the complex concepts of quantum effects.

First trials with graduate and undergraduate students have shown good indications of the educational value of the game. 
Already after a couple of trial games with voluntary physics students, the understanding of quantum operations have considerably improved.
The students not only understood how the states were evolving but they also conceived and tested better strategies to achieve the highest possible qudit values to win the game.
The predominantly positive feedback shows promise for enhanced student involvement in future quantum information courses. 
An enthusiastic student, for example, stated that he ``learned about quantum logic in an engaging, fun way.'' 
However, we note that a thorough study to evaluate the educational effectiveness of the game would be needed.

\section{Limitations and future improvements}
The current version of the game describes some fundamental quantum mechanical effects, where it focuses highly on quantum computation. 
Other interesting quantum mechanical effects, such as a continuous wave-like probability distribution beyond the equally-weighted superposition of qudit states, are beyond the scope of this game.

Although the first trial games have given positive feedback overall, the game can be further improved. 
An additional operation that could be introduced to the game is a state measurement operation which measures the state of one or more qudits, individually, at any point in the quantum circuit. 
This mechanic would add an extra layer of complexity and would allow the game to introduce simple quantum algorithms, such as superdense coding, quantum teleportation, or entanglement swapping.

\section{Conclusion}

Teaching quantum mechanical concepts is a challenging task, not only because students usually have an obstructive perception of quantum mechanics, but also because they often lack an intuitive comprehension. 
To loosen up the atmosphere and promote student engagement, methods from gamification can be applied. 
The presented strategic card game \emph{Endless Fun in high dimensions} offers multiple game modes with which learning new quantum computational concepts is facilitated and diversified.
Additionally, the underlying fundamental quantum features, namely superpositions, interference, and entanglement, can be experienced and understood in a quantum computing setting.
First trial games with students have shown the educational effectiveness, and potential the game has for supporting conventional teaching methods. 
Together with the evaluation software, the card game is a powerful tool which is not only suitable for players with background knowledge but also for introducing players to quantum operations in an easy-going way. 
Thus, it can also be used for outreach purposes where interested laymen can experience fundamental quantum physical features and the functioning of a quantum computer.

\begin{acknowledgments}
The authors thank Stephen Plachta, Matias Eriksson, Subhajit Bej, Marco Ornigotti, Mona Pulst, and Rupa Kiran for feedback on the game, valuable suggestions, and design support. 
The authors furthermore thank Ilkka Kylänpää for help in software-related questions.
The authors acknowledge the inventors of the card game \emph{Q$\ket{\text{Cards}}$}: Oskari Kerppo, Jorden Senior, Sabrina Maniscallo, Guillermo Garcia-Perez, Samuli Jääskeläinen, Sylvia Smatanova, Krista Erkkilä, and Elie Abraham. 
All authors acknowledge financial support from the Academy of Finland through the Competitive Funding to Strengthen University Research Profiles (decision 301820) and the Photonics Research and Innovation Flagship (PREIN - decision 320165). 
LK acknowledges support from the Vilho, Yrjö and Kalle Väisälä Foundation of the Finnish Academy of Science and Letters.
MH acknowledges support from from the Doctoral School of Tampere University and the Magnus Ehrnrooth foundation through its graduate student scholarship. 
RF acknowledges support from the Academy of Finland through the Academy Research Fellowship (decision 332399).
\end{acknowledgments}

\section*{Conflict of Interest Disclosure}
The authors have no conflicts to disclose.

\newpage
\newpage
\,


\section*{Supplementary material}

# Endless Fun

By the Experimental Quantum Optics Group, Tampere University, Finland

## The Idea of the game

Build your own high-dimensional quantum computer. By applying quantum logic operations to high-dimensional quantum bits – so-called *qudits* -, you can change their values. Try to reach the highest possible value on your qudit, while keeping the other player's qudit as low as possible. The evaluation software calculates the final state after each round, in which a certain number of measurements will project possible complex superposition states into a final value. The player with the highest qudit values summed up from all three rounds wins the game.

## What you will need

- 1-7 players
- 55 Cards (You can find the exact number of the different cards in the "Cards" section of this manual)
- As many pawns as players
- 1 die
- Paper and pen to write down the points
- The evaluation software

## Setup

- The players agree on which version to play:
    - **Easy version**: This is an easy method to get started. It is played in 2d, i.e. qubit values of 0 and 1 are possible, and without the phase properties. Take the following cards out of the card deck: Z, Y, H2, H3.
    - **2D version:** Qubit values of 0 and 1 are possible. In 2d, qubit values of 0 and 1 are possible. The 2d version adds cards that change the phase of single states which can control quantum interference. The H3 cards are removed from the card deck.
    - **3D version:** The qudit values can be 0, 1 and 2. All cards are used.
- Every player gets a pawn and 7 random cards, which the players hide from each other.
- Decide on a starting player by throwing a dice. The player with the highest number starts.
- The starting player is the first to choose his or her qudit position. The player marks the qudit's position by placing his or her pawn.
- Then, the other players take turns placing their pawns (in clockwise direction from starting player). The players can place their pawns only on positions neighboring already set pawns.

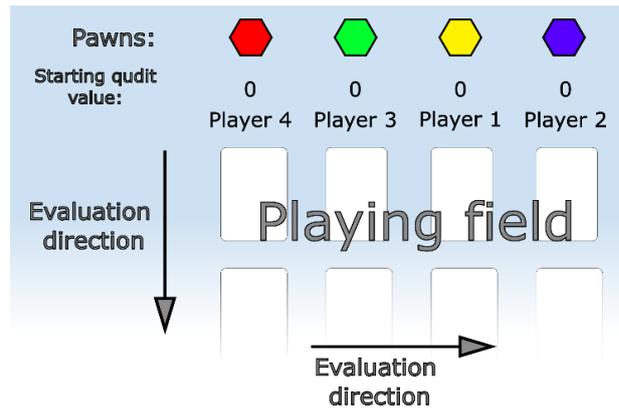

The layout at the start of a 4-player game after the starting positions are decided.

# Playing the game

### 1st Round:
- The starting qudit value for all players is 0.
- The starting player starts the round by playing one card from his or her hand.
- Going clockwise, the other players take turn in playing their operations.
- The round is finished after every player has played three cards.
- Then, the played quantum operations are entered into the evaluation software to determine the winning state.
- The players earn as many points as the value of their qudit in the round's winning state. Write the points down on a paper.

### 2nd Round:
Round 2 is played as the previous round with the following changes:
- The starting state of all players is their winning value of the last round.
- The starting player is the player to the left side of the previous starting player.
- Every player plays only 2 cards.

### 3rd Round:
Round 3 is played as the previous round with the following changes:
- Only 1 card is played per player.

### End of the game:
- The player with the most points summed up over all three rounds wins the game.

## Cooperative mode

The cooperative mode is a game mode in which the focus is not set on individual achievement, as it is in the main game mode. In this mode, the same rules apply as in the main competitive mode. However, the goal is not to win against the other players but to reach the highest possible values summed up over all qudits as a group. The players are not allowed to show their cards to the other players and cannot discuss strategies.

## Single-player mode

In the single-player mode, players can solve riddles given in the additional material of the game. The instructions for each riddle are given on the riddle cards. Therefore, the regular setup and gameplay of the multi-player mode do not apply in this mode, but the rules on how to play cards and enter the operations in the evaluation software must be followed. The solutions along with some background information can be found on the back of the cards. The players can use the evaluation software for help.

## How to play cards

There are three different types of quantum operations. For all quantum operations the cards must be placed on the highest possible position beneath a qudit, but spaces to the left or right are allowed. Take a look at the truth tables at the end of the instructions to learn how they act.

> **Note:**
> - The players can put the card on any position, i.e. also apply logic operations on qudits of other players.
> - The way a computation on this quantum computer works: every unitary is applied from left to right sequentially and top to bottom

1. **Single qudit operations**

   Single qudit cards act only on one qudit and are placed beneath the qudit on which they should act on.

   **The cards:**

   | | | |
   |---|---|---|
   | 4x | I | Identity operation. It does nothing. |
   | 4x | X | $X_1$ and $X_2$ operation in 3d; X operation in 2d |
   | 4x | Y | $Y_1$ and $Y_2$ operation in 3d; Y operation in 2d |
   | 4x | Z | $Z_1$ and $Z_2$ operation in 3d; Z operation in 2d |
   | 15x | Hadamard | $H_1$, $H_2$ and $H_3$ operation in 3d; $H_1$ and $H_2$ operation in 2d |

A possible way to place cards and the respective evaluation in 2d could be:

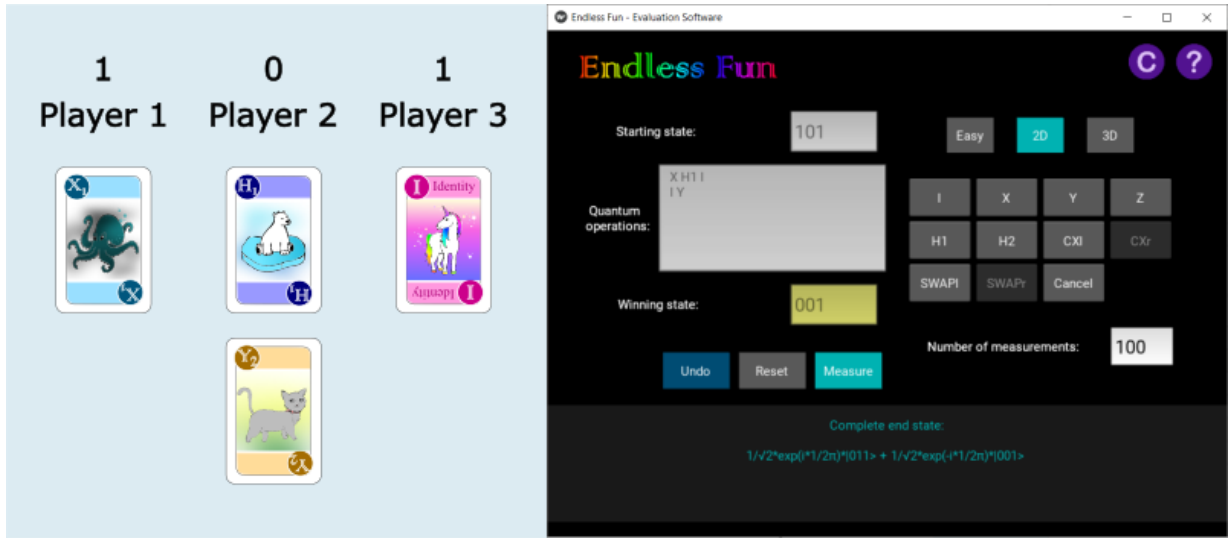

It is not allowed to leave empty spaces beneath a qudit.

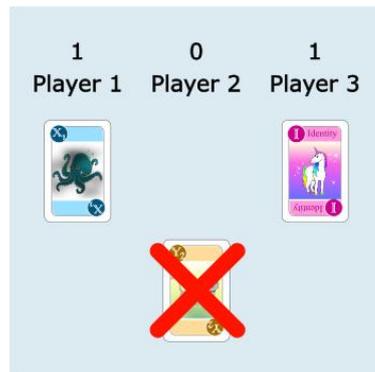

## 2. Two qudit operations

These cards act on two quantum states. The quantum operations can only be played on neighboring qudits. Tilting the card signifies on which neighboring state the quantum operation acts on.

| **The cards:** | | |
|---|---|---|
| 9x | CX | CX operation. The chosen control and target are placed next to each other on the field. The card is placed on the control qudit and tilted to the target qudit. In the software a CX with control on the left and target on the right is entered as CXr (and vice versa). |
| 8x | Swap | Swaps the state with the chosen neighboring state. A tilt of the card signifies which states are swapped. In the software, swapping a state with the neighboring state on the right is entered as SWAPr (and vice versa). |

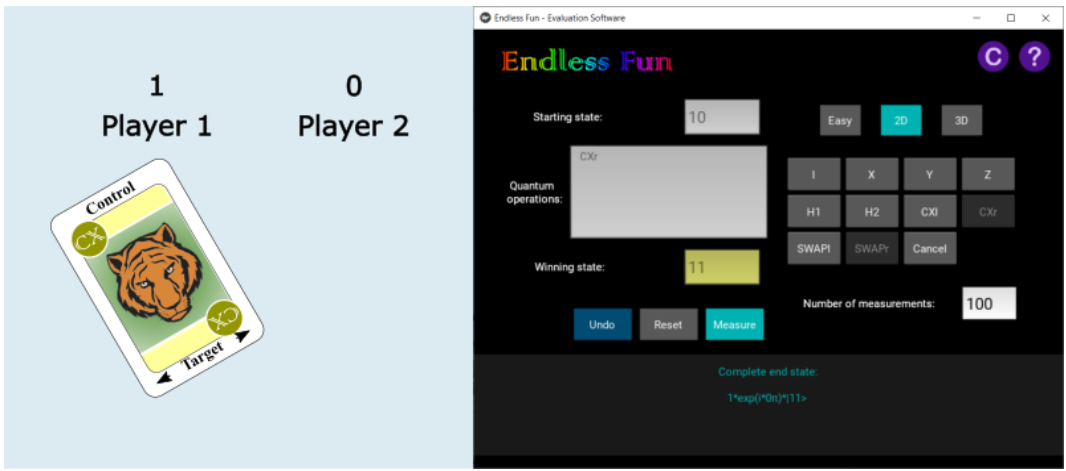

Here, the state of player 1 acts as the control and the state of player 2 as the target of the CX gate.

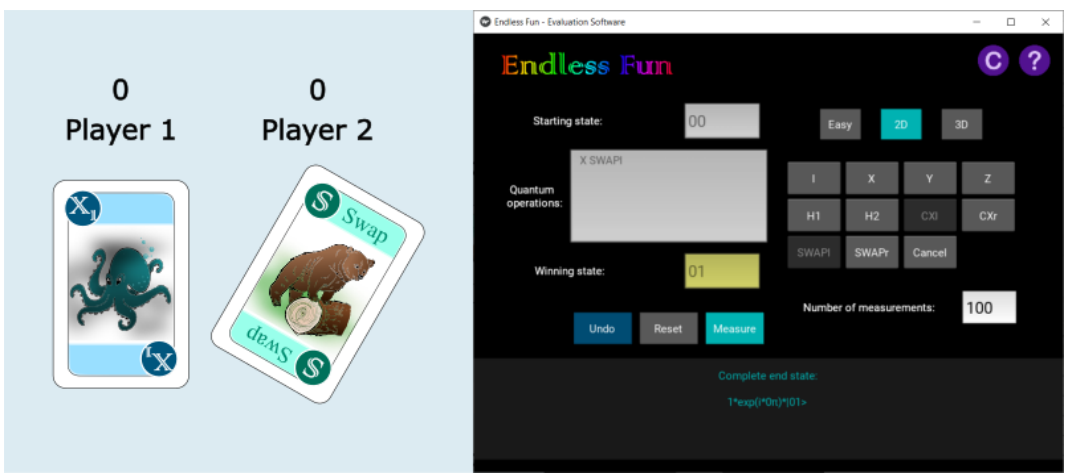

The swap card is tilted into the direction of the state, with which the qudit states should be swapped.

## 3. Other cards

### The cards:

| | | |
|---|---|---|
| 2x | Steal | Draw a random card from someone else and play it immediately. |
| 5x | Cancel | Cancel the operation of any card already played. They subsequently turn into identities. |

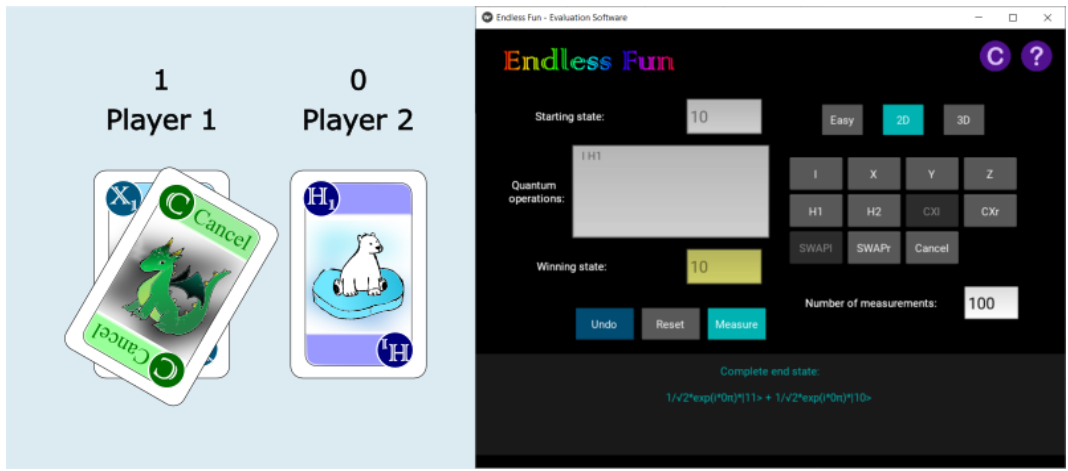

The cancel card is placed on top of the card that should be nullified. It is then entered as an identity operation in the software.

## The evaluation software

- Run the evaluation application. Running the software in python is more powerful.
- Choose if you want to play in the easy version, in 2d, or 3d by clicking on the buttons in the top right.
- Input the starting state
  - In 2d qubit values of 0 and 1 are allowed. In 3d the possible qudit values are 0, 1 and 2.
- Input the played quantum logic gates, by clicking on the buttons in the right panel.
  - Remember that the quantum computer applies the operations sequentially from left to right and top to bottom. The gates must also be entered in that order.
  - Use the identity operation as a placeholder for empty spaces.
- Define the number of measurements or use the default value of 100.
- Click the 'Measure'-button to compute the winning state. The winning state is shown in the winning state text field, which is highlighted in yellow once the state is calculated.
  - If the Measure button is not enabled, some inputs are incorrect. Click the reset button next to it and enter the parameters again.
  - The winning state is the state, which was statistically measured the most often in the measurements.
- Click on the 'Reset'-button to evaluate the next round.

**Note:**
- The more measurements, the smaller the deviation of the outcome from the theoretical probability is, i.e. less luck is involved. In other words: one measurement leads to a pretty random result.
- The complete state is displayed at the bottom. It is the quantum state before its projection onto a single state through a measurement.

A first round played with 4 players in 3d in which player 1 starts with a qudit value of 0, player 2 with 2, player 3 with 1 and player 4 with 0 could for example be:

Since the winning state is 2111, player 1 earns 2 and the other players 1 point in this round.

# Truth tables of the quantum gates

### Identity

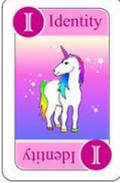

**3D**
$|0\rangle \to |0\rangle$
$|1\rangle \to |1\rangle$
$|2\rangle \to |2\rangle$

**2D**
$|0\rangle \to |0\rangle$
$|1\rangle \to |1\rangle$

### X-gate

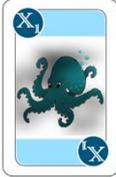

**3D**
$|0\rangle \to |1\rangle$
$|1\rangle \to |2\rangle$
$|2\rangle \to |0\rangle$

$|0\rangle \to |2\rangle$
$|1\rangle \to |0\rangle$
$|2\rangle \to |1\rangle$

**2D**
$|0\rangle \to |1\rangle$
$|1\rangle \to |0\rangle$

---

### Y-gate

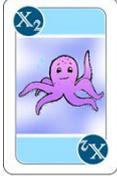  $Y_1 = Z_1 X_1$

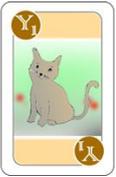  $Y_2 = Z_2 X_2$

**3D**
$|0\rangle \to e^{i\frac{2\pi}{3}} |1\rangle$
$|1\rangle \to e^{-i\frac{2\pi}{3}} |2\rangle$
$|2\rangle \to |0\rangle$

$|0\rangle \to e^{i\frac{2\pi}{3}} |2\rangle$
$|1\rangle \to |0\rangle$
$|2\rangle \to e^{-i\frac{2\pi}{3}} |1\rangle$

**2D**
$|0\rangle \to i|1\rangle$
$|1\rangle \to i|0\rangle$

### Z-gate

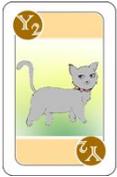

**3D**
$|0\rangle \to |0\rangle$
$|1\rangle \to e^{i\frac{2\pi}{3}} |1\rangle$
$|2\rangle \to e^{-i\frac{2\pi}{3}} |2\rangle$

$|0\rangle \to |0\rangle$
$|1\rangle \to e^{-i\frac{2\pi}{3}} |1\rangle$
$|2\rangle \to e^{i\frac{2\pi}{3}} |2\rangle$

**2D**
$|0\rangle \to |0\rangle$
$|1\rangle \to -|1\rangle$

---

### H-gate
Hadamard

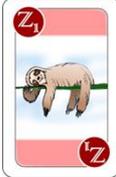 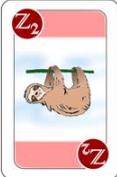 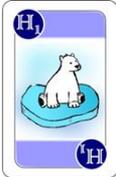

**3D**

$|0\rangle \to \frac{1}{\sqrt{3}}\left(|0\rangle + e^{i\frac{2\pi}{3}}|1\rangle + e^{i\frac{2\pi}{3}}|2\rangle\right)$
$|1\rangle \to \frac{1}{\sqrt{3}}\left(|0\rangle + e^{-i\frac{2\pi}{3}}|1\rangle + |2\rangle\right)$
$|2\rangle \to \frac{1}{\sqrt{3}}\left(|0\rangle + |1\rangle + e^{-i\frac{2\pi}{3}}|2\rangle\right)$

$|0\rangle \to \frac{1}{\sqrt{3}}\left(|0\rangle + e^{-i\frac{2\pi}{3}}|1\rangle + e^{-i\frac{2\pi}{3}}|2\rangle\right)$
$|1\rangle \to \frac{1}{\sqrt{3}}\left(|0\rangle + |1\rangle + e^{i\frac{2\pi}{3}}|2\rangle\right)$
$|2\rangle \to \frac{1}{\sqrt{3}}\left(|0\rangle + e^{i\frac{2\pi}{3}}|1\rangle + |2\rangle\right)$

$|0\rangle \to \frac{1}{\sqrt{3}}(|0\rangle + |1\rangle + |2\rangle)$
$|1\rangle \to \frac{1}{\sqrt{3}}\left(|0\rangle + e^{i\frac{2\pi}{3}}|1\rangle + e^{-i\frac{2\pi}{3}}|2\rangle\right)$
$|2\rangle \to \frac{1}{\sqrt{3}}\left(|0\rangle + e^{-i\frac{2\pi}{3}}|1\rangle + e^{i\frac{2\pi}{3}}|2\rangle\right)$

**2D**
$|0\rangle \to |0\rangle + |1\rangle$
$|1\rangle \to |0\rangle - |1\rangle$

$|0\rangle \to |0\rangle + i|1\rangle$
$|1\rangle \to |0\rangle - i|1\rangle$

|   |   | CX | | SWAP | |
|---|---|---|---|---|---|
| | | 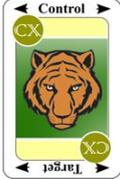 | | | 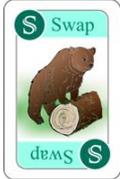 |

| | CX | | SWAP | |
|---|---|---|---|---|
| **3D** | $\|0\rangle\|0\rangle \to \|0\rangle\|0\rangle$<br>$\|0\rangle\|1\rangle \to \|0\rangle\|1\rangle$<br>$\|0\rangle\|2\rangle \to \|0\rangle\|2\rangle$<br>$\|1\rangle\|0\rangle \to \|1\rangle\|1\rangle$<br>$\|1\rangle\|1\rangle \to \|1\rangle\|2\rangle$<br>$\|1\rangle\|2\rangle \to \|1\rangle\|0\rangle$<br>$\|2\rangle\|0\rangle \to \|2\rangle\|2\rangle$<br>$\|2\rangle\|1\rangle \to \|2\rangle\|0\rangle$<br>$\|2\rangle\|2\rangle \to \|2\rangle\|1\rangle$ | **3D** | $\|0\rangle\|0\rangle \to \|0\rangle\|0\rangle$<br>$\|0\rangle\|1\rangle \to \|1\rangle\|0\rangle$<br>$\|0\rangle\|2\rangle \to \|2\rangle\|0\rangle$<br>$\|1\rangle\|0\rangle \to \|0\rangle\|1\rangle$<br>$\|1\rangle\|1\rangle \to \|1\rangle\|1\rangle$<br>$\|1\rangle\|2\rangle \to \|2\rangle\|1\rangle$<br>$\|2\rangle\|0\rangle \to \|0\rangle\|2\rangle$<br>$\|2\rangle\|1\rangle \to \|1\rangle\|2\rangle$<br>$\|2\rangle\|2\rangle \to \|2\rangle\|2\rangle$ | |
| **2D** | $\|0\rangle\|0\rangle \to \|0\rangle\|0\rangle$<br>$\|0\rangle\|1\rangle \to \|0\rangle\|1\rangle$<br>$\|1\rangle\|0\rangle \to \|1\rangle\|1\rangle$<br>$\|1\rangle\|1\rangle \to \|1\rangle\|0\rangle$ | **2D** | $\|0\rangle\|0\rangle \to \|0\rangle\|0\rangle$<br>$\|0\rangle\|1\rangle \to \|1\rangle\|0\rangle$<br>$\|1\rangle\|0\rangle \to \|0\rangle\|1\rangle$<br>$\|1\rangle\|1\rangle \to \|1\rangle\|1\rangle$ | |

# Credits

NOTE: The general idea is strongly based on the great card game *Q|Cards>* by Oskari Kerppo, Jorden Senior, Sabrina Maniscallo, Guillermo Garcia-Perez, Samuli Jääskeläinen, Sylvia Smatanova, Krista Erkkilä, Elie Abraham, which is a very similar game using only qubits. It can be downloaded here:
https://zhamul.itch.io/qcards

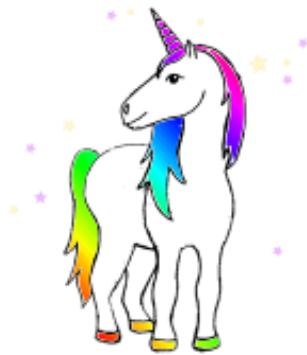

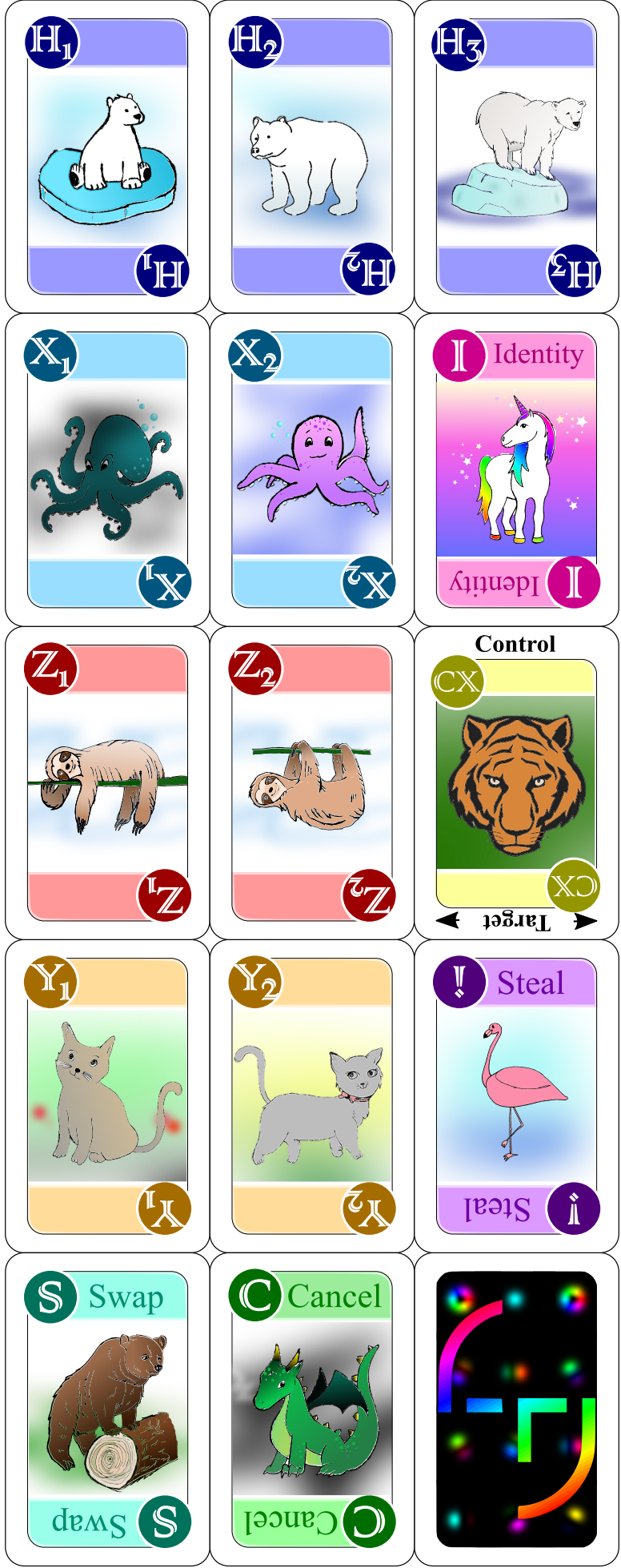

Number of needed cards:

4x I 9x CX
4x X 8x Swap
4x Y 2x Steal
4x Z 5x Cancel
15x H

# Bit Flip

**Easy**

You have one qubit with value 0 in **2D**.

Flip the value to 1 without using an X- or Y-gate.

# Bit Flip - Solution

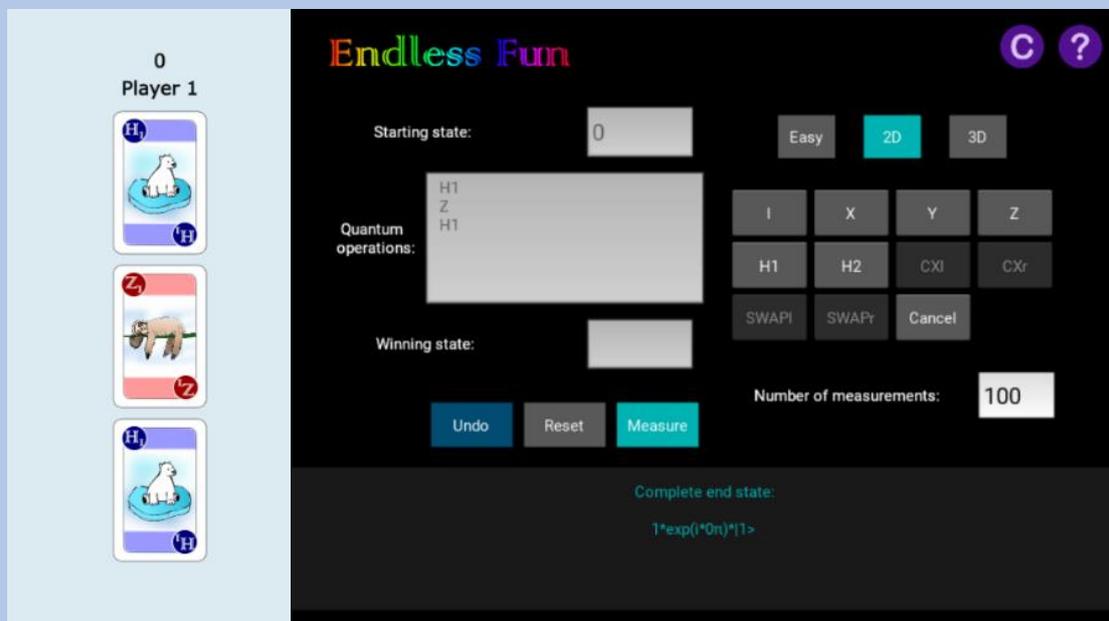

**Background:** Quantum interference is one of the most important concepts in quantum computing. It is at the heart of almost all quantum enhancements, such as finding a solution to a specific problem exponentially faster than a classical computer. Two of the most popular enhanced quantum algorithms are factorizing prime numbers (Shor algorithm) and the search through a large unstructured data set (Grover algorithm).

# Dream big

**Easy**

You start with a $|\psi\rangle = |2\rangle + |1\rangle + |0\rangle$ quantum state in **3D**.
(Play an $H_1$ gate on a $|0\rangle$ input state to get the initial state.)

Make sure that you will win the round with 2 points by generating the state $|2\rangle$ without playing more than three cards.
Don't use any X or Y gates.

# Dream big - Solution

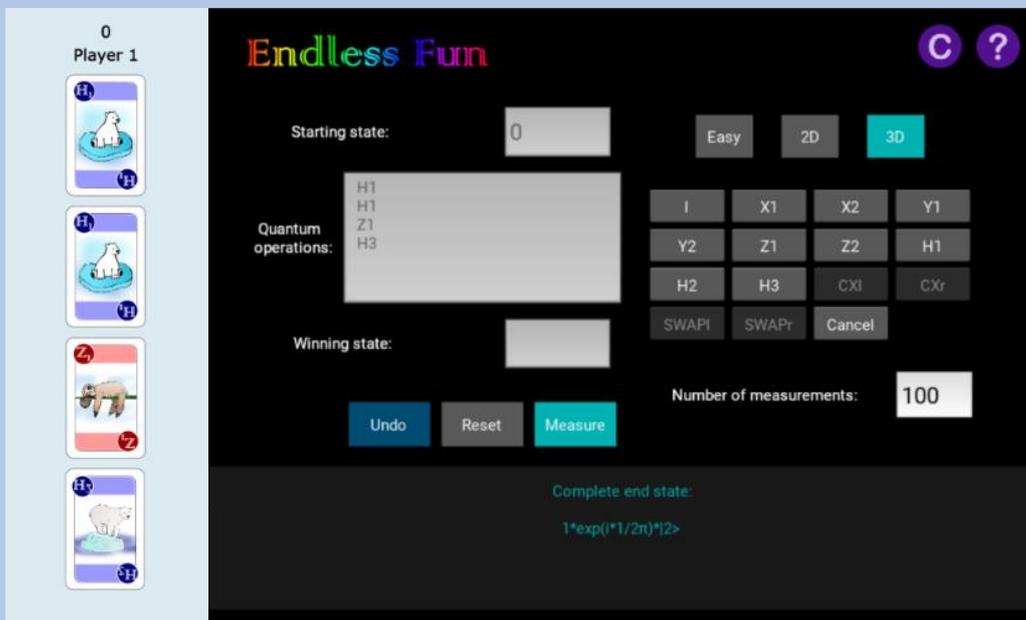

**Note to the solution:** Interference with qudits requires an additional Hadamard gate in contrast to the 2D case (see "Bit-Flip" riddle).

**Background:** Although the field of high-dimensional quantum information is still in its infancy, it is a promising candidate for future quantum computers. It is already known, e.g., that qudits can simplify some quantum computational tasks as they require less multi-qudit operations.

# Let's Swap

You have two arbitrary qubits in **2D**.

Swap the two qubit values whatever their input is.
You can't use a SWAP-gate.

# Let's Swap - Solution

**Background:** All quantum computational tasks can be performed using only basic quantum logic gates, which are called elementary quantum gates. By cleverly arranging them, it is possible to perform complex operations and large-scale computational tasks. Building a SWAP gate out of CX gates is a simple example of it.

# Well-Balanced Odds

**Advanced**

You start with a |000⟩ quantum state in **3D**.

Play quantum operations so that there is only one unpredictable winner each round.

# Well-Balanced Odds - Solution

**Background:** Entanglement can not only be realized between two qubits, but also between several qudits. Entangled states between many quantum systems are an integral part in quantum computing and are jointly responsible for the quantum-computational speed-up. They are also very fragile: one broken qudit destroys the entanglement between all.

# Jingle Bells

**Expert**

Start with the state |00⟩ in **2D**.

Generate a quantum state with only one unpredictable winner. This condition should still hold if you play $H_1$ or $H_2$ gates on both states at the same time.

# Jingle Bells - Solution

**Background:** The generated state is the anti-symmetric $|\Psi^-\rangle$ Bell state. Like all Bell-states, $|\Psi^-\rangle$ is maximally entangled, i.e. the local outcome is entirely random while the global outcome is perfectly correlated. $|\Psi^-\rangle$ is the only state that always gives anti-correlated outcomes independent of the measurement direction. H-gates change the measurement direction and toggle all other Bell states between correlated and anti-correlated states.

# The Third Wheel

**Expert**

You start with a |000⟩ quantum state in **3D**.

Create a state in which the first and last qudit are entangled in three dimensions. The qudit in the middle should be uncorrelated.

*Hint: Generate a maximally entangled state (such as a GHZ state) first. Then disentangle the second qudit.*

# The Third Wheel - Solution

**Note to the solution:** The second qudit can only have one value ($|0\rangle$), hence it is separable from the other two qudits, which remain entangled.

**Background:** The architecture of quantum computers does not always allow operations on distant qudits. Therefore, it is important to have the possibility to generate links, such as entanglement here, between distant nodes even if they can't directly interact with each other.


\end{document}